\documentclass[preprint,12pt]{elsarticle}

\usepackage{epsfig}
\usepackage{enumerate}

\usepackage{amssymb,amsmath,graphics,graphicx}  
\usepackage[dvipsnames]{xcolor}

\journal{Elsevier}

\begin{document}

\begin{frontmatter}


\title{A mathematical model for the bullying dynamics in schools}

\author{Nuno Crokidakis}
\ead{nunocrokidakis@id.uff.br}

\address{Instituto de F\'{\i}sica, Universidade Federal Fluminense, Niter\'oi, Rio de Janeiro, Brazil}

\begin{abstract}
We analyze a mathematical model to understand the dynamics of bullying in schools. The model considers a population divided into four groups: susceptible individuals, bullies, individuals exposed to bullying, and violent individuals. Transitions between these states occur at rates designed to capture the complex interactions among students, influenced by factors such as romantic rejection, conflicts with peers and teachers, and other school-related challenges. These interactions can escalate into bullying and violent behavior. The model also incorporates the role of parents and school administrators in mitigating bullying through intervention strategies. The results suggest that bullying can be effectively controlled if anti-bullying programs implemented by schools are sufficiently robust. Additionally, the conditions under which bullying persists are explored.

\end{abstract}


\end{frontmatter}



\section{Introduction}

The dynamics of social systems has attracted the attention of scientists across various disciplines, ranging from social to natural sciences. From a mathematical perspective, the approach involves translating sociological concepts into variables, which are then analyzed using system dynamics techniques. This process results in the construction of a set of differential and algebraic equations that describe the evolution of these variables over time \cite{bronson,galam_book,socio_review}. 

A social phenomenon that has been minimally explored from a mathematical perspective is bullying. It is defined as long-term, repeated victimization, where an individual or a group seeks to harm, intimidate, or coerce someone perceived as vulnerable \cite{olweus}. Bullying occurs when people repeatedly and intentionally use words or actions against an individual or group to cause distress and jeopardize their well-being \cite{human_rights}.

A recent study analyzed the impact of bullying on future juvenile crimes \cite{sourander}. According to the findings, frequent bullies and individuals who both bullied and were bullied (comprising $8.8\%$ of the sample) accounted for $33.0\%$ of all juvenile crimes over the 4-year study period. Being frequently a bully alone was associated with both occasional and repeated offending, while a bully-victim status specifically predicted repeated offending. Bullying was found to be a predictor for various types of crimes—including violence, property offenses, drunk driving, and traffic violations—even when controlling for parental education level \cite{sourander}.

An important question is: how can bullying in schools be effectively controlled? School leaders and teachers play a crucial role in preventing and addressing bullying. Bullying often occurs in classrooms and hallways, where teachers and administrators are in a position to intervene. Additionally, classrooms serve as an ideal environment for educating children about bullying and its consequences. It is equally important for schools to take responsibility for addressing bullying in accordance with established rules and policies. In this context, another significant question arises: how should parents collaborate with schools if their children are bullying others? It is suggested that parents engage with the school to determine appropriate, proportional consequences for their child. Discipline should always be immediate, non-violent, and focused on correcting behavior and fostering rehabilitation, rather than humiliation or punitive measures \cite{unicef}.

Bullying does not occur exclusively in classrooms. Teachers and administrators need to recognize that bullying often takes place in less supervised areas, such as bathrooms, playgrounds, crowded hallways, and school buses, as well as through digital platforms like cell phones and computers, where supervision is minimal or absent. Despite these challenges, bullying must always be taken seriously. Educators should emphasize that reporting bullying is not the same as tattling. If a teacher witnesses bullying in the classroom, they must intervene immediately to stop it, document the incident, and report it to the appropriate school administrators to ensure a proper investigation is conducted \cite{apa}.

The above-mentioned works address the problem primarily from a qualitative perspective, such as through discussions with students, teachers, and school administrators, and by analyzing the results of these conversations or surveys. To the best of our knowledge, no mathematical models have been developed to address this issue. The study of complex social dynamics, such as the spread of bullying, is indeed a challenging task. However, simple mathematical models can provide valuable insights into various social systems \cite{galam_fake,Ibrahim,sooknanan,sooknanan2,celia_new,andre,maj_vote_pmco,pmco_jstat,lucas,monteiro,francheschi,javarone,nuno,chuang,wang}.

In this work, we propose a compartmental model to study the evolution of bullying dynamics in schools. Initially, we analyze a homogeneous mixing population by formulating a set of ordinary differential equations that describe the behavior of each subpopulation. After the analysis of such equations, we conduct numerical simulations of the model on two-dimensional grids to explore the impact of bullying dynamics in smaller, localized populations. Our results, both analytical and numerical, indicate that bullying can be mitigated or even eradicated under specific conditions. These conditions, determined by the parameters defined in the model, are examined in detail in the subsequent sections.


\section{The Model}

Each individual in our artificial population can be in one of four possible states: (1) susceptible individuals (S), who have never been bullied, or who were bullied in the past but are not currently experiencing bullying, or individuals who engaged in bullying in the past but have since stopped; (2) bullies (B), individuals who actively practice bullying; (3) exposed individuals (E), who are victims of bullying but do not engage in violent actions against bullies; and (4) violent individuals (V), who are bullied and respond with violent behavior (e.g., aggression) \cite{sommer,aydin}. A schematic representation of the transitions between states $S, B, E$ and $V$ is shown in Fig. 1.

\begin{figure}[t]
\begin{center}
\vspace{6mm}
\includegraphics[width=0.7\textwidth,angle=0]{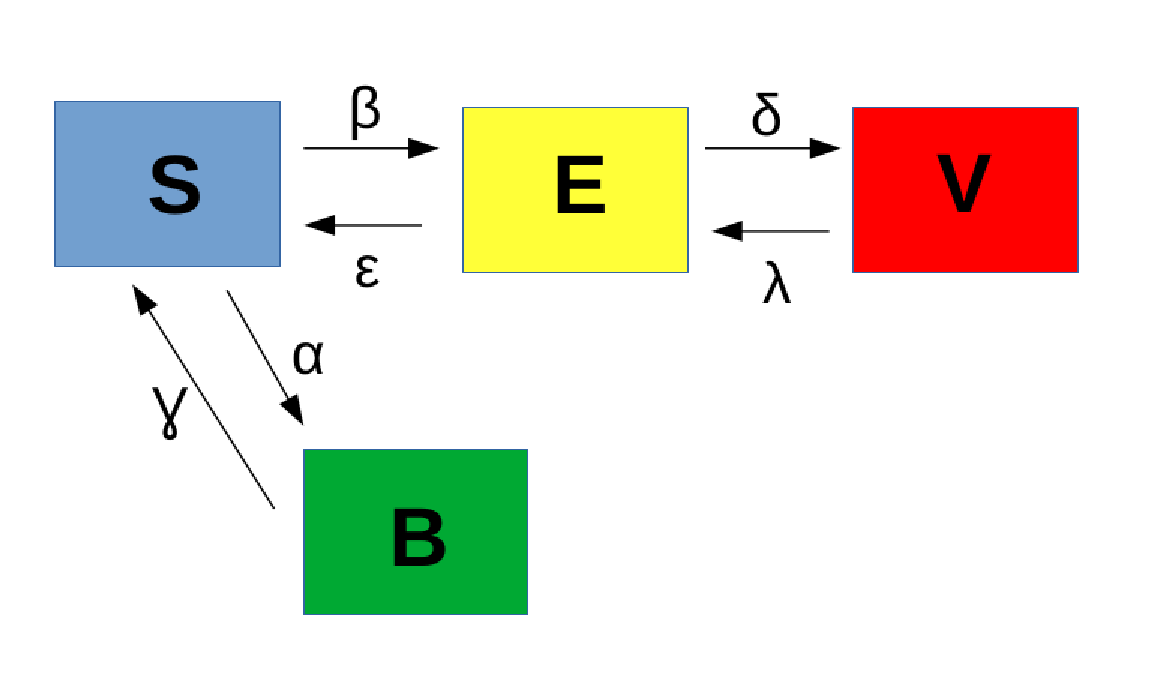}
\end{center}
\caption{(Color online) Schematic representation of the model's transitions, showing the four possible states:  Susceptible individuals (\textbf{S}), Bully individuals (\textbf{B}), Exposed individuals (\textbf{E}) and Violent individuals (\textbf{V}). The parameters responsible for each transition are also shown.}
\label{fig1}
\end{figure}

In Fig. \ref{fig1} , we observe that the transition from $S$ to $B$ occurs with probability $\alpha$, but it requires a social interaction between $S$ and $B$ agents. The transition from $S$ to $E$ can occur with probability $\beta$ if the $S$ individual interacts with a $B$ individual. The remaining transitions between states are spontaneous, meaning they do not require social interactions between distinct individuals: (i) $E\to S$ occurs with probability $\epsilon$; (ii) $B\to S$ occurs with probability $\gamma$;  (iii) $E\to V$ occurs with probability $\delta$; and (iv) $V\to E$ occurs with probability $\lambda$.

The interpretation of the above parameters is as follows. The parameter $\alpha$ represents a contagion probability, where $S$ individuals become bullies $B$. It reflects the social pressure that lead individuals who neither suffer nor practice bullying to engage in bullying behavior. The parameter $\beta$ represents the probability of interaction between bullying individuals $B$ and those who do not experience bullying $S$, leading $S$ individuals to become exposed $E$ to bullying. This transition models scenarios such as social rejection based on factors like race, weight, or height, where $S$ individuals start to experience bullying \cite{sommer}. Individuals in state $E$ have not yet responded with violence against the bullying. Indeed, as discussed in \cite{sommer}, as most students experience rejection, even those who are bullied and ostracized by peers or teachers do not resort to violence. We also introduce the parameter $\epsilon$, which represents the probability that an exposed individual (E) stops experiencing bullying. This can occur, for example, if the families of bullying individuals intervene by discussing the consequences of bullying behavior with them \cite{unicef}. In such cases, the $E$ individual may return to state S (see further discussion in the next paragraph). The parameter $\gamma$ is the probability that a bullying individual $B$ stops bullying and transitions back to state $S$. This can be seen as a measure of school control over bullying behavior, modeling the impact of anti-bullying school programs. The parameter $\delta$ represents the probability of an exposed student $E$, who is suffering bullying, becoming a violent individual $V$ as a result. Violent individuals can also stop engaging in violent acts, for example, due to parental influence or school intervention. In such cases, they return to the exposed state $E$ with probability $\lambda$, since they may not necessarily stop suffering bullying. We can also consider other transitions between distinct states, such as $S \to V$. However, as discussed earlier, not all students who are victims of bullying engage in violent behavior \cite{sommer}. Moreover, including additional transitions would require the introduction of new parameters, and we aim to keep the model as simple as possible.

As discussed in \cite{menestrel}, similar to individuals who only engage in bullying, individuals who both bully and are bullied often exhibit heightened aggression compared to their non-involved peers. To simplify the model and avoid considering a direct transition from $E\to  B$ or introducing another class of individuals who both bully and are bullied simultaneously, we assume that exposed $E$ individuals can return to the susceptible $S$ state. After this, some of these individuals may transition to the bully $B$ state. This reflects the idea that not all individuals who are bullied will go on to bully others. Therefore, some individuals who undergo the transition $E\to S$ may later become bullies $B$, while others may not.

In addition, some studies suggest that victims of bullying may try to speak with their parents or school management. However, due to the lack of action to stop the bullying or out of fear of speaking to others, many victims feel compelled to take matters into their own hands \cite{usp}. This justifies the inclusion of the V state in the model.

In recent years, epidemic-like compartmental models have been widely used to study various social dynamics (a recent review of these models can be found in \cite{sooknanan}). These models are particularly useful because they treat behaviors as potentially contagious, offering a simple yet powerful framework for understanding a wide range of social phenomena, such as crime, opinions, addiction, and fanaticism \cite{sooknanan}. In the present manuscript, we extend this approach to study bullying dynamics in schools. Our model conceptualizes bullying as a form of social contagion, where the spread of aggressive behavior occurs through interactions among students, teachers, and school administrators. By incorporating probabilities that account for the possibility of non-contagion, the model provides a nuanced view of how bullying propagates and offers a novel perspective on this complex phenomenon.

In the next section, we present and discuss the results of our model. We begin by analyzing a system of ordinary differential equations under the assumption of a homogeneous mixing population. Subsequently, we explore the outcomes of simulations performed on two-dimensional grids, providing a more spatially explicit perspective.


\section{Results}

\subsection{Analytical Framework: Ordinary Differential Equations for Bullying Dynamics}

Let us represent the time-dependent populations of Suceptible, Bully, Exposed and Violent individuals as $S=S(t), B=B(t), E=E(t)$ and $V=V(t)$, respectively. Thus, one can define the densities $s=S/N, b=B/N$, $e=E/N$ and $v=V/N$, where $N$ is the population size. Assuming a homogeneously mixing population and applying the interaction rules outlined in Section 2, the model is governed by the following set of coupled differential equations:
\begin{eqnarray} \label{eq7}
\frac{ds}{dt} & = & -\beta\,s\,b - \alpha\,s\,b +\gamma\,b + \epsilon\,e \\  \label{eq8}
\frac{db}{dt} & = & \alpha\,s\,b - \gamma\,b \\ \label{eq9}
\frac{de}{dt} & = & \beta\,s\,b - \delta\,e + \lambda\,v - \epsilon\,e \\ \label{eq10}
\frac{dv}{dt} & = & \delta\,e - \lambda\,v
\end{eqnarray}

\noindent

We can now discuss the basic reproduction number  $R_o$ of the model. This number quantifies the expected number of cases directly generated by one individual in a population where all individuals are susceptible to the phenomenon \cite{anderson_may}. It is commonly defined in epidemic spreading models. In our case, the ``infected'' individual would be one who is practicing bullying at the initial time $t=0$. Thus, we can consider the conditions for $t=0$ as $s(t=0)=1-1/N, b(t=0)=1/N$ and $e(t=0)=v(t=0)=0$. In such a case, Eq. (\ref{eq8}) can be approximated at initial times to \cite{allen}
\begin{equation} \label{eq12}
\frac{db}{dt} = (\alpha-\gamma)\,b  ~,
\end{equation}  
\noindent
that is valid for large populations, where $s(0)\approx 1$. After integration, Eq. \eqref{eq12} gives us $b(t)=b^{'}\,e^{\gamma\,(R_o-1)\,t}$. In this last result, $b^{'}$ is a contant, and $R_o$ is given by
\begin{equation} \label{eq13}
R_o = \frac{\alpha}{\gamma} ~.
\end{equation}
\noindent
The results given by Eqs. \eqref{eq12} and \eqref{eq13} are valid for initial times. The rapid increase of bully individuals, i.e., the outbreak in the language of dynamics of epidemic spreading \cite{allen}, will occurr if $\alpha > \gamma$, which gives $R_o > 1$. In other words, for the initial time evolution of the population, the contagion probability $\alpha$ and the spontaneous transition probability $\gamma$ drive the dynamics. We will discuss more about this point in the following.

\begin{figure}[t]
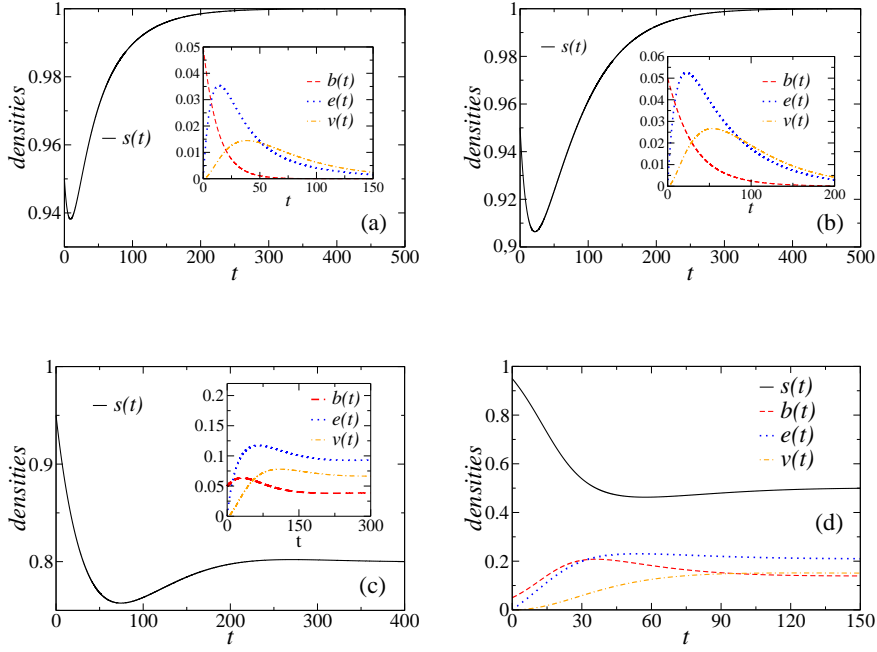

\begin{center}
\vspace{6mm}
\includegraphics[width=0.4\textwidth,angle=0]{figure2a.eps}
\hspace{0.3cm}
\includegraphics[width=0.4\textwidth,angle=0]{figure2b.eps}
\\
\vspace{1.0cm}
\includegraphics[width=0.4\textwidth,angle=0]{figure2c.eps}
\hspace{0.3cm}
\includegraphics[width=0.4\textwidth,angle=0]{figure2d.eps}
\end{center}
\caption{Time evolution of the four densities of agents $s(t), b(t), e(t)$ and $v(t)$, obtained from the numerical integration of Eqs. (\ref{eq7}) - (\ref{eq10}). The parameters are $\beta=0.30, \gamma=0.20, \delta=0.05, \lambda=0.07$ and $\epsilon=0.10$, and we varied $\alpha$: (a) $\alpha=0.05$, (b) $\alpha=0.15$, (c) $\alpha=0.25$ and (d) $\alpha=0.40$. All parameters are given in unities year$^{-1}$. For the considered parameters, we have from Eq. (\ref{eq13}): (a) $R_o = 0.25$, (b) $R_o = 0.75$, (c) $R_o = 1.25$, (d) $R_o = 2.00$.}
\label{fig2}
\end{figure}

Before analyzing the equilibrium solutions of the model, we can discuss about the evolution of the populations $S, B, E,$ and $V$ in time. We can obtain the evolution of the densities $s(t), b(t), e(t)$ and $v(t)$ through numerical integration of Eqs. \eqref{eq7} - \eqref{eq10}. To visualize the evolution of the subpopulation densities $s(t), b(t), e(t)$ and $v(t)$, we exhibit in Fig. \ref{fig2} the time evolution of those quantities. The initial conditions are $s(0)=0.95, b(0)=0.05$ and $e(0)=i(0)=0.0$. The parameters are $\beta=0.30, \gamma=0.20, \delta=0.05, \lambda=0.07$ and $\epsilon=0.10$, and we varied the contagion probability $\alpha$. In Fig. \ref{fig2} (a) we exhibit results for $\alpha=0.05$. In such a case, we do not expect an outbreak in the curve $b(t)$. Indeed, we observe that $b(t)\to 0$ exponentially, but we can see some increase of the curves for exposed $e(t)$ and violent $v(t)$ individuals. Since the Susceptible agents decrease slowly in time, due to the fast disappearence of Bully individuals, the subpopulations of Exposed and Violent individuals appears. However, this phenomenon does not last long: the subpopulations $b(t), e(t)$ and $v(t)$ disappear after some time, and the population becomes composed of only Susceptible agents.

In Fig. \ref{fig2} (b) we observe a similar phenomena for $\alpha=0.15$. However, due to a higher contagion probability $\alpha$, the subpopulation $s(t)$ decreases to lower values in comparison with the case $\alpha=0.05$. Due to this lower values, the subpopulations $e(t)$ and $i(t)$ achieve higher values in the initial times. When we increase $\alpha$, we increase the contagion probability leading to the transition $S\to B$, leading to more bully individuals, which will also lead to the increase of the compartments $E$ and $V$. We also exhibit in Fig. \ref{fig2} results for $\alpha=0.25$ (panel (c)) and $\alpha=0.40$ (panel (d)). For such values, we observe the outbreak in the curves for $b(t)$, and the long-time survival of the subpopulations $b(t), e(t)$ and $v(t)$. The equilibrium values can be found analytically, as we will discuss in the following.

We can obtain the equilibrium or seady-state solutions of Eqs. \eqref{eq7} - \eqref{eq10} (see the Appendix). For these equations, we have have two distinct solutions, and the first one is a bullying-free (\textit{bf}) equilibrium where only susceptible agents remain in the population. This solution is given by  
\begin{equation}\label{eq14}
\{s_{bf}, b_{bf}, e_{bf}, v_{bf}\}  = \{1,0,0,0\} ~,
\end{equation}
\noindent
where the index $bf$ denotes the bullying-free equilibrium solution. On the other hand, the another solution is an endemic one, namely
\begin{equation} \label{eq15}
\{s_{end}, b_{end}, e_{end}, v_{end}\}  = \left\{\frac{\gamma}{\alpha}, b_{end}, \frac{\beta\,\gamma}{\epsilon\,\alpha}\,b_{end}, \frac{\beta\,\gamma\,\delta}{\epsilon\,\alpha\,\lambda}\,b_{end}\right\} ~,
\end{equation}
\noindent
where the index $end$ denotes the endemic equilibrium solution, and $b_{end}$ is the endemic density of bully individuals, 
\begin{equation} \label{eq16}
b_{end} = \frac{\epsilon\,\lambda\,(\alpha-\gamma)}{\epsilon\,\alpha\,\lambda + \beta\,\gamma\,(\lambda+\delta)} ~.
\end{equation}

As pointed before, Eq. \eqref{eq14} represents a bullying-free equilibrium. In other words, the bullying can be controlled in certain conditions, that will be discussed in the following. On the other hand, Eq. \eqref{eq15} represents an endemic solution, i.e., bullying remains in the population. Looking for Eq. \eqref{eq16}, we can see that the endemic solution only makes sense for $\alpha>\gamma$, since a population density cannot be negative. Thus, for $\alpha>\gamma$ we have the solution given by Eq. \eqref{eq15} as the valid solution for the model, whereas for $\alpha<\gamma$ we have the solution given by Eq. \eqref{eq14}. From Eq. \eqref{eq16}, we observe that $b_{end}$ increases with $\lambda$; this indicates that when the number of students practicing violet acts decreases, it can stimulate a new wave of bullying pratices. We can also see that $b_{end}$ decreases with $\delta$. In other words, when the violence of individuals that suffer bullying grows, it can inhibit the practice of bullying in schools \cite{neto}. In addition, the rise of anti-bullying scholl programmes (increasing $\gamma$) leads to the decrease of the equilibrium solution $b_{end}$ \cite{anti-bullying}. Also, $b_{end}$ increases with $\alpha$ (the recruitment parameter of susceptible individuals to start to practice bullying). Notice also that the endemic solution for the density of violent individuals, $v_{end}$, increases with $b_{end}$. In other words, the more students practicing bullying, the more violent actions committed by students that suffer bullying.

\begin{figure}[t]
\begin{center}
\vspace{6mm}
\includegraphics[width=0.7\textwidth,angle=0]{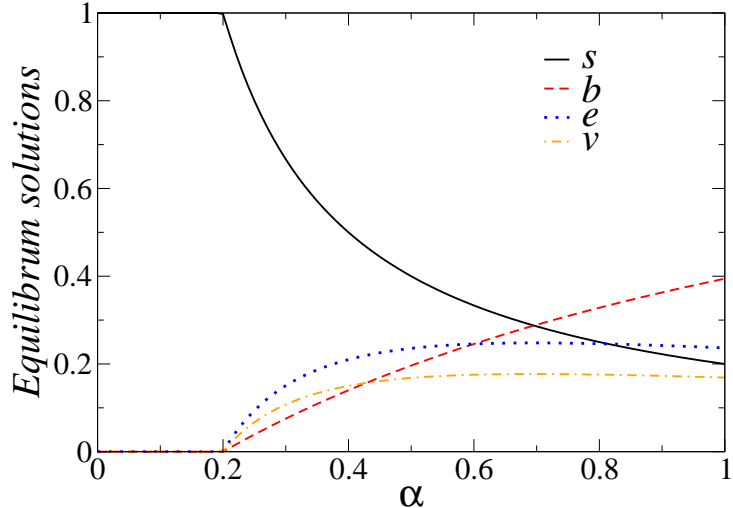}
\end{center}
\caption{(Color online) Equilibrium values $s$ (full line), $b$ (dashed line), $e$ (dotted line) and $v$ (dotted-dashed line) as functions of $\alpha$. The lines were obtained from the numerical integration of Eqs. \eqref{eq7} - \eqref{eq10}. Both solutions, bullying-free and endemic ones, given by Eqs. \eqref{eq14} and \eqref{eq15}, respectively, are shown in the figure. The parameters are $\beta=0.30, \gamma=0.20, \delta=0.05, \lambda=0.07$ and $\epsilon=0.10$. It is important to note that these behaviors are present also for other parameter values, and what is shown here works as a pattern.}
\label{fig3}
\end{figure}

In Fig. \ref{fig3} we present the equlibrium solutions  $s, b, e$  and $v$ as functions of the contagion probability $\alpha$. The fixed parameters are $\beta=0.30, \gamma=0.20, \delta=0.05, \lambda=0.07$ and $\epsilon=0.10$. The curves were generated from the numerical integration of the model's equations \eqref{eq7} - \eqref{eq10}, and agree very well with the analytical results given by Eqs. \eqref{eq14} and \eqref{eq15}. Particularly, we observe the bullying-free equilibrium solution given by Eq. \eqref{eq14} for $\alpha<\gamma=0.20$, whereas for $\alpha>\gamma=0.20$ we observe the endemic solution given by Eq. \eqref{eq15}.

The above discussion regarding the validity of the equilibrium solutions can be formalized by analyzing the stability of the stationary solutions given by Eqs. \eqref{eq14} and \eqref{eq15}. The stability analysis can be performed considering the Jacobian matrix $J$ of the model's equations \eqref{eq7} - \eqref{eq10}, 
\begin{equation} \label{eq177}
J = 
\begin{bmatrix}
-(\alpha+\beta)\,b & -(\alpha+\beta)\,s+\gamma & \epsilon & 0 \\
\alpha\,b          & \alpha\,s-\gamma          & 0        & 0 \\ 
\beta\,b           & \beta\,s                  & -(\delta+\epsilon) & \lambda \\
0                  & 0                         & \delta       & -\lambda
\end{bmatrix}
\end{equation}

For the matrix given by Eq. \eqref{eq177}, we found a fourth-order polynomial for the eigenvalues $\Lambda$ of the form $\Lambda^{4} + A\,\Lambda^{3} + B\,\Lambda^{2} + C\,\Lambda + D = 0$. Thus, we need to consider separately the two distinct stationary solutions of the model. Starting from the bullying-free equilibrium, given by Eq. \eqref{eq14}, we have the following four eigenvalues:
\begin{eqnarray} \label{eq17}
\Lambda_1 & = & 0 \\  \label{eq18}
\Lambda_2 & = & \alpha-\gamma \\  \label{eq19}
\Lambda_3 & = & -\delta -\epsilon \\  \label{eq20}
\Lambda_4 & = & -\lambda 
\end{eqnarray}
\noindent
Following the Routh-Hurwitz criterion, an equilibrium point is locally asymptotically stable if all eigenvalues of $J$ have negative real parts \cite{rw}. Remembering that all model's parameters are probabilities per unit time, all of them are positive, which directly leads to $\Lambda_3<0$ and $\Lambda_4<0$. On the other hand, the eingenvalue $\Lambda_2$ is negative for $\alpha<\gamma$, which agrees with the condition we obtained above for the validity of the solution given by Eq. \eqref{eq14}. Notice also that $\Lambda_2$ can be rewriten as $\Lambda_2=\gamma\,(R_o-1)$, that is negative for $R_o<1$.

In the case of the endemic solution, Eq. \eqref{eq15}, the above-mentioned fourth-order polynomial has the following coefficients,
\begin{eqnarray}
A & = & \lambda + \delta + \epsilon + (\alpha+\beta)\,\epsilon\,\lambda\,\alpha\,\Theta \\
B & = & \lambda(\epsilon+\delta) + (\alpha+\beta)\,\epsilon\,\lambda\,\alpha\,(\epsilon+\delta+\lambda)\Theta  \\
C & = & \lambda(\epsilon+\delta)\,(\alpha+\beta)\,\epsilon\,\lambda\,\alpha\,\Theta  \\
D & = & \epsilon\,\lambda^{2}\,\alpha\,\beta\,\gamma\,\delta\,\Theta
\end{eqnarray}  
\noindent
with 
\begin{equation}
\Theta = \frac{1}{\epsilon\,\alpha\,\lambda + \beta\,\gamma\,(\lambda+\delta)}\,\left(1-\frac{1}{R_o}\right) ~,
\end{equation}
\noindent
and again $R_o$ is given by Eq. \eqref{eq13}. Thus, considering the Routh–Hurwitz criteria for local stability \cite{rw}, all above coefficients need to be positive, and one can see that the endemic solution is locally asymptotically stable for $R_o>1$, i.e., for $\alpha>\gamma$, as obtained above.



\subsection{Simulation Framework: Bullying Dynamics on Grids}

We also considered the model on two-dimensional grids of size $L$ x $L$. For this purpose, we performed numerical simulations of the model considering the rules defined in section 2. Thus, we built an agent-based formulation of the compartmental model proposed in section 2, and the algorithm to simulate the model is defined as follows:

\begin{itemize}

\item we generate a $L$ x $L$ grid or square lattice with a population size $N=L^{2}$ and periodic boundary conditions;

\item given an initial condition $S(0), B(0), E(0)$ and $V(0)$, we randomly distribute these agents in the lattice sites;

\item at each time step, every lattice site is visited in a sequential order;

\item if a given agent $i$ is in $S$ state, we choose at random one of his/her four nearest neighbors, say $j$. If such neighbor $j$ is in $B$ state, agent $i$ changes to $E$ state with probability $\beta$ or change to $B$ state with probability $\alpha$.

\item if a given agent $i$ is in $B$ state, he/she changes to state $S$ with probability $\gamma$;

\item if a given agent $i$ is in $E$ state, he/she changes to $V$ state with probability $\delta$ or change to $S$ state with probability $\epsilon$.

\item if a given agent $i$ is in $V$ state, he/she changes to $E$ state with probability $\lambda$.

\end{itemize}

One time step is defined by the visit of all lattice sites. The agents' states were updated syncronously, i.e., we considered parallel updating, as it is standard in probabilistic cellular automata in order to avoid correlations between consecutive steps \cite{roos,religion}.

\begin{figure}[t]
\begin{center}
\vspace{6mm}
\includegraphics[width=0.7\textwidth,angle=0]{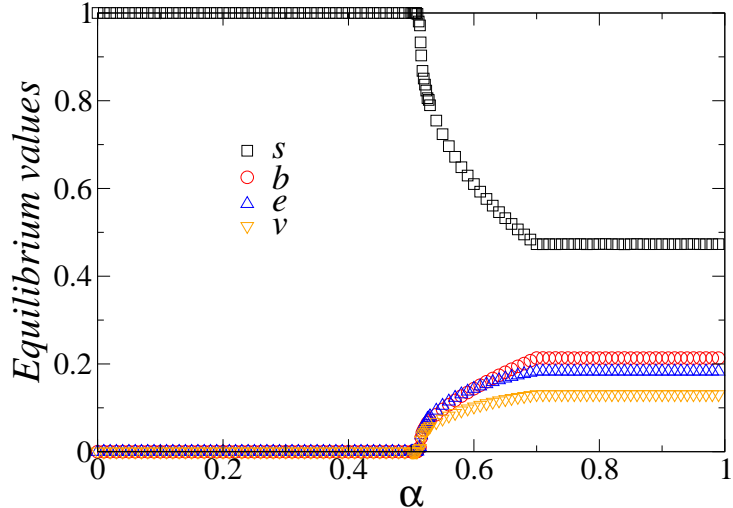}
\end{center}
\caption{(Color online) Equilibrium values $s$ (squares), $b$ (circles), $e$ (up triangles) and $v$ (down triangles) as functions of $\alpha$ obtained from numerical simulations of the model on a grid of linear size $L=60$. The parameters are $\beta=0.30, \gamma=0.20, \delta=0.05, \lambda=0.07$ and $\epsilon=0.10$. Results are averaged over $100$ independent simulations.}
\label{fig4}
\end{figure}

The time evolution of the densities $s(t), b(t), e(t)$ and $v(t)$ are qualitatively similar to we observe in the fully-connected (mean field) graph case (not shown). As we can see in Fig. \ref{fig4}, the equilibrium values obtained from the simulations are also qualitatively similar to the previous case, but the value of the parameter $\alpha$ at which we observe a change in the solutions (from endemic to bullying-free equilibrium) occurs at $\alpha\approx 0.5$ for the same parameters considered in the homogeneous mixing case, where we observed the limit of validity of the solutions at $\alpha=\gamma=0.2$. For the numerical results, we considered a grid of linear size $L=60$.

\begin{figure}[t]
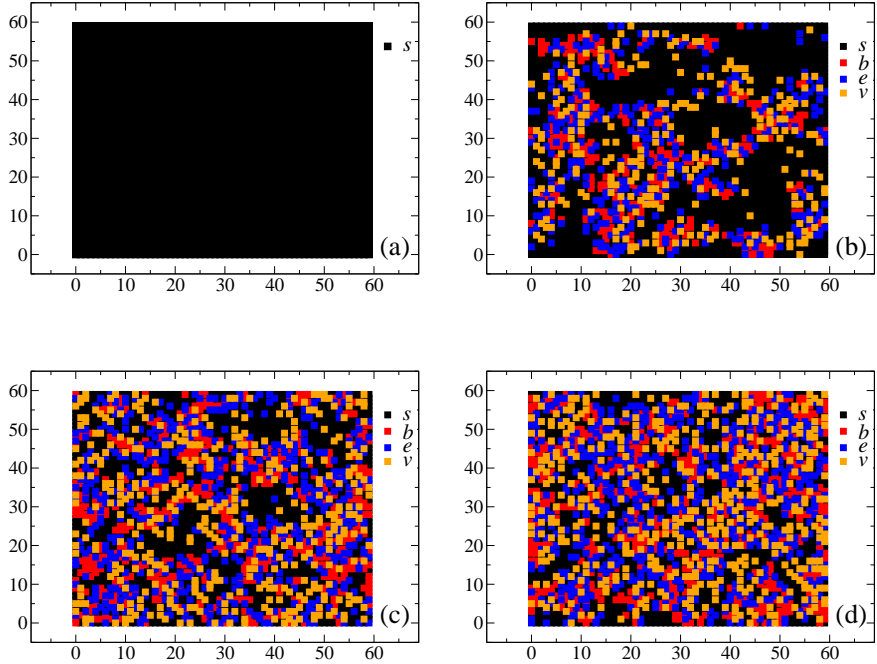

\begin{center}
\vspace{6mm}
\includegraphics[width=0.4\textwidth,angle=0]{figure5a.eps}
\hspace{0.3cm}
\includegraphics[width=0.4\textwidth,angle=0]{figure5b.eps}
\\
\vspace{1.0cm}
\includegraphics[width=0.4\textwidth,angle=0]{figure5c.eps}
\hspace{0.3cm}
\includegraphics[width=0.4\textwidth,angle=0]{figure5d.eps}
\end{center}
\caption{Snapshots of the population on a grid of linear size $L=60$ at stationary states. The fixed parameters are $\beta=0.30, \gamma=0.20, \delta=0.05, \lambda=0.07$ and $\epsilon=0.10$, and we varied $\alpha$: (a) $\alpha=0.45$, (b) $\alpha=0.55$, (c) $\alpha=0.65$ and (d) $\alpha=0.85$. The squares' colors represent distinct subpopulations, namely $S$ (black), $B$ (red), $E$ (blue) and $V$ (orange).}
\label{fig5}
\end{figure}

For better visualization of the grid and the subpopulations, we plot in Fig. \ref{fig5} some snapshots of the steady states of the model. For this figure we considered the grid size as $L=60$, the same fixed parameters $\beta=0.30, \gamma=0.20, \delta=0.05, \lambda=0.07$ and $\epsilon=0.10$, and we plot four distinct values of $\alpha$. The distinct colors represent the subpopulations  $S$ (black), $B$ (red), $E$ (blue) and $V$ (orange). For smaller values of $\alpha$, such as $\alpha=0.45$, we observe the survival of only the Susceptible population, i.e., the subpopulations $B, E$ and $V$ disappear from the population after a long time (panel (a)). When $\alpha$ is increased to $\alpha=0.55$, we observe that the subpopulations $B, E$ and $V$ survive and coexist with the Susceptible population, but the Susceptibles are still the majority (panel (b)). For larger values of $\alpha$, such as $\alpha=0.65$ (panel (c)) or $\alpha=0.85$ (panel (d)), the Susceptible subpopulation decreases, and we observe an increase in the other three compartments $B, E$ and $V$.


\section{Final Remarks}   

\quad In this work, we present a mathematical model to study the spread of bullying in schools. The model is composed of four classes of agents: susceptible individuals ($S$), bully individuals ($B$), exposed individuals ($E$), and violent individuals ($V$). These individuals can change states due to social interactions, but we also consider state changes that occur spontaneously, independent of such interactions.

The impact of these distinct types of state changes was discussed. In particular, the analytical and numerical results concerning the model's rate equations show the existence of a locally stable bullying-free equilibrium, where only susceptible individuals remain in the equilibrium states. On the other hand, there is also an endemic equilibrium where bullying activities continue to occur in the school, driven by bullying individuals, and all four states $, B, E$ and $V$ are present in the population. Bullying can be effectively controlled if anti-bullying school programs and parental actions are effective in educating children. The results suggest that these actions should be improved in both schools and families in order to better control the bullying problem in schools. Numerical results of the model defined on two-dimensional lattices were also discussed.


\section*{Acknowledgments}

The author acknowledges partial financial support from the Brazilian scientific funding agency Conselho Nacional de Desenvolvimento Cient\'ifico e Tecnol\'ogico (CNPq, Grant 308643/2023-2).

\appendix
\section{Analytical considerations}

In this appendix we will detail some of the analytical calculations to obtain the equilibrium solutions of the model.

We will consider Eqs. \eqref{eq7} - \eqref{eq10} in the long-time limit ($t\to\infty$), and the normalization condition given by $s+b+e+v=1$. From Eq. \eqref{eq8}, we obtain two solutions, namely $b=0$, which represents the bullying-free (\textit{bf}) equilibrium, and $s=\gamma/\alpha$, that is valid for $b\neq 0$.

From Eq. \eqref{eq9}, we have
\begin{equation} \label{ap1}
b=\frac{\epsilon\,e}{\beta\,s}
\end{equation}
\noindent
For the \textit{bf} equilibrium ($b=0$), Eq. \eqref{ap1} leads to $e=0$. Considering Eq. \eqref{eq10}, one can obtain the relation
\begin{equation} \label{ap2}
v=\frac{\delta\,e}{\lambda} ~,
\end{equation}
\noindent
leading to $v=0$ in the limit of validity of the \textit{bf} equilibrium ($\alpha<\gamma$, as discussed in the text). Thus, when the \textit{bf} equilibrium is valid, we have $b=e=v=0$, and the normalization conditions leads to $s=1$. This represents the bullying-free solution given by Eq. \eqref{eq14}.

For the endemic solution, we have to consider the previous obtained result for the density of susceptible individuals, namely $s=\gamma/\alpha$. Considering this result, and Eq. \eqref{ap2} rewritten as $e=\lambda\,v/\delta$, plugging these two results in Eq. \eqref{ap1} leads to
\begin{equation}\label{ap3}
b = \frac{\epsilon\,\lambda\,\alpha}{\beta\,\gamma\,\delta}\,v
\end{equation}

Thus, with the last results, the normalization conditions gives us
\begin{equation} \label{ap4}
v=\frac{\beta\,\gamma\,\delta}{\epsilon\,\alpha\,\lambda+\beta\,\gamma\,(\lambda+\delta)}\left(1-\frac{\gamma}{\alpha}\right)
\end{equation}

Considering Eq. \eqref{ap4}, Eq. \eqref{ap3} can be rewritten as
\begin{equation} \label{ap5}
b_{end}=\frac{\epsilon\,\lambda\,(\alpha-\gamma)}{\epsilon\,\alpha\,\lambda + \beta\,\gamma\,(\lambda+\delta)} ~.
\end{equation}
\noindent
This last result is Eq. \eqref{eq16} of the text. In addition, from Eq. \eqref{ap3} we have the endemic solution for $v$ writen as function of $b$, namely
\begin{equation}\label{ap6}
v_{end}=\frac{\beta\,\gamma\,\delta}{\epsilon\,\alpha\,\lambda}\,b_{end}
\end{equation}
The endemic solution for $e$ can be obtained considering Eq. \eqref{ap6} in Eq. \eqref{ap2}, leading to
\begin{equation}\label{ap7}
e_{end} = \frac{\beta\,\gamma}{\epsilon\,\alpha}\,b_{end}
\end{equation}

Thus, considering the result $s_{end}=\gamma/\alpha$ and Eqs. \eqref{ap5}, \eqref{ap6} and \eqref{ap7}, we have the endemic solution of the model.

\vspace{0.8cm}

\bibliographystyle{elsarticle-num-names}

\end{document}